\documentclass[prx,twocolumn,english,superscriptaddress,floatfix,longbibliography]{revtex4-2}

\usepackage{graphicx}% Include figure files
\usepackage{dcolumn}% Align table columns on decimal point
\usepackage{bm}% bold math

\usepackage[separate-uncertainty=true]{siunitx}

\usepackage{physics}

\begin{document}

\preprint{APS/123-QED}

\title{Degenerate Photons from a Cryogenic Spontaneous \\Parametric Down-Conversion Source}

\author{Nina Amelie Lange}
\email{nina.amelie.lange@upb.de}
\author{Timon Schapeler}
\author{Jan Philipp H\"opker}
\author{Maximilian Protte}
\author{Tim J. Bartley}

\affiliation{Department of Physics \& Institute for Photonic Quantum Systems, Paderborn University, Warburger Str. 100, 33098 Paderborn, Germany}

\date{\today}% It is always \today, today,

\begin{abstract}
We demonstrate the generation of degenerate photon pairs from spontaneous parametric down-conversion in titanium in-diffused waveguides in lithium niobate at cryogenic temperatures. Since the phase-matching cannot be temperature tuned inside a cryostat, we rely on a precise empirical model of the refractive indices when fabricating a fixed poling period. We design the phase-matching properties of our periodic poling to enable signal and idler photons at (1559.3~$\boldsymbol{\pm}$~0.6)~nm, and characterize the indistinguishability of our photons by performing a Hong-Ou-Mandel interference measurement. Despite the effects of photorefraction and pyroelectricity, which can locally alter the phase-matching, we achieve cryogenic indistinguishable photons within \SI{1.5}{nm} to our design wavelength. Our results verify sufficient understanding and control of the cryogenic nonlinear process, which has wider implications when combining quasi-phase-matched nonlinear optical processes with other cryogenic photonic quantum technologies, such as superconducting detectors.
\end{abstract}

\maketitle

\section{Introduction}
Spontaneous parametric down-conversion (SPDC) under ambient conditions is a well-established technique for the generation of quantum light, such as heralded single photons~\cite{hong1986experimental}, entangled photon pairs~\cite{kwiat1995new}, and squeezed states~\cite{wu1986generation}. Integrated SPDC in particular benefits from high conversion efficiencies and the propagation of photon pairs into well-defined waveguide modes~\cite{fujii2007bright,alibart2016quantum}. This allows for the combination of integrated SPDC with other components to manipulate and detect the photons on a single chip~\cite{kim2020hybrid,wang2020integrated,moody20222022,pelucchi2022potential}. In order to set up a fully integrated circuit including an SPDC source, two major challenges need to be overcome. On the one hand, integrated pump suppression together with integrated detection remains challenging~\cite{harris2014integrated,gentry2018monolithic}. 
On the other hand, all components must be compatible in their operating conditions~\cite{wang2020integrated}. 

In this work, we address the second issue by adapting SPDC to the challenging operating conditions of superconducting detectors. 
Superconducting nanowire single-photon detectors (SNSPDs) provide detection efficiencies close to unity~\cite{you2020superconducting,reddy2020superconducting,chang2021detecting}, low dark-count rates~\cite{chiles2022new}, and high timing resolution~\cite{korzh2020demonstration}. However, they must be operated below their critical temperature, and therefore require cryogenic operation~\cite{elshaari2020hybrid}. To combine integrated SPDC together with high performance SNSPDs, cryogenic SPDC becomes indispensable. Moreover, understanding how nonlinear processes can be designed to operate under cryogenic conditions is critical not just for nonlinear quantum light sources, but also for frequency conversion of other cryogenic photonic technologies~\cite{bartnick2021cryogenic}.

We have recently shown the proof-of-principle functionality of cryogenic integrated type-II SPDC in titanium in-diffused lithium niobate waveguides~\cite{Lange2022cryogenic}. While our SPDC source remained fully functional after changing the temperature by two orders of magnitude, there was a large shift in the phase-matching properties. Understanding the temperature-dependent changes in the SPDC process is important to predict and customize the nonlinear interaction at cryogenic temperatures. Here we demonstrate that we can achieve cryogenic degeneracy by tailoring the phase-matching properties of the nonlinear waveguide. We optimize the poling period to overlap the wavelengths of signal and idler in the telecom C-band. We demonstrate the indistinguishability by characterizing the joint spectral intensity (JSI) and performing a Hong-Ou-Mandel (HOM) interference measurement.

\section{Temperature dependence of SPDC}
The spectral properties of signal and idler are determined by energy and momentum conservation, in practice described by the spectral distribution of the pump beam and the phase-matching function~\cite{grice2001eliminating}. In general, the desired signal and idler wavelengths are often not phase-matched because of the dispersion of the nonlinear crystal.
Efficient interaction of a particular wavelength combination can be achieved by periodic inversion of the spontaneous polarization, also known as quasi-phase-matching~\cite{tanzilli2001highly}. While the pump distribution is independent of the waveguide temperature, quasi-phase-matching relies on the crystal dispersion, the length of the poled region, and poling period. All three parameters are temperature dependent and need to be considered when investigating signal and idler properties under cryogenic operation. 

Degenerate SPDC can be understood as the reverse process of classical second harmonic generation (SHG)~\cite{couteau2018spontaneous}. 
As a preliminary characterization, we thus start by performing an SHG measurement to determine the phase-matched wavelength, which is equal to the degenerate signal and idler wavelength.
The generated SHG power $P_{\mathrm{SHG}}$ at temperature $T$ scales with the phase-mismatch of the interacting fields $\Delta k$ and the effective length of the poled region $L$ as~\cite{boyd2008nonlinear}
\begin{equation}\label{eqn:shgPower}
P_{\mathrm{SHG}}(T) \propto \mathrm{sinc}^2 \left( \frac{\Delta k(T) L(T)}{2} \right) .
\end{equation}
The phase-mismatch for a type-II process can be described by 
\begin{equation}\label{eqn:mismatch}
\Delta k(T) = 2\pi \left( \frac{\Delta n(\lambda_\mathrm{p},T)}{\lambda_\mathrm{p}}  - \frac{1}{\Lambda(T)} \right) ,
\end{equation}
with 
$\Delta n(\lambda_\mathrm{p},T) = 2n_{\mathrm{TE}}(\lambda_\mathrm{p}/2,T) - n_{\mathrm{TE}}(\lambda_\mathrm{p},T) - n_{\mathrm{TM}}(\lambda_\mathrm{p},T)$ combining the temperature dependent effective refractive indices $n$ of the TE- or TM-polarized modes, pump wavelength $\lambda_\mathrm{p}$, and poling period $\Lambda$. Under ambient conditions, the temperature dependence is used to tune the phase-matching, but this method is incompatible with cryogenic operation. Therefore, we require precise knowledge of the desired poling period at low temperatures, which is fixed during fabrication.

We can describe our waveguide dispersion by applying Sellmeier equations for bulk lithium niobate~\cite{edwards1984temperature,jundt1997temperature}, which are extrapolated for temperatures below room temperature. Furthermore, we include the estimated increase in the effective refractive indices due to the titanium in-diffusion, to account for our waveguide geometry. This is done by using a commercial mode solving software (RSoft FemSIM), which is based on the finite element method (for details, see~\cite{bartnick2021cryogenic}). The change in the length of the poled region and poling period are given by the thermal contraction of lithium niobate, for which empirical data for temperatures down to \SI{60}{K} are available~\cite{scott1989properties}. For temperatures below \SI{60}{K}, we assume the length to be unchanged, since the thermal expansion coefficient goes to zero at \SI{0}{K}.

\begin{figure}[t]
\centering\includegraphics[width=0.92\linewidth]{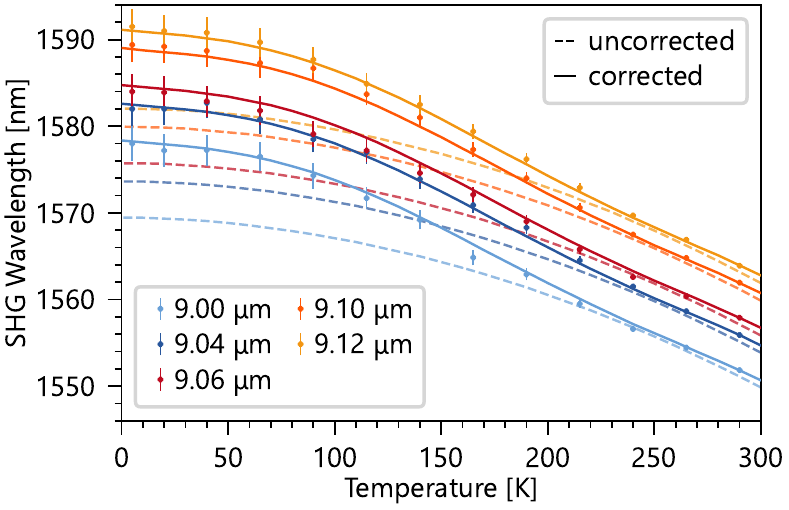}
\caption{\label{fig:SHGShift} Shift in the temperature-dependent phase-matched SHG wavelength for five different poling periods measured in a step-wise cool-down. The dashed lines show the uncorrected simulation and the solid lines represent the corrected simulation which includes the empirical correction term.}
\end{figure}

The extrapolation of the effective refractive indices introduces an uncertainty of our simulations. We address this problem by applying an additional experimentally determined correction to the refractive index data, using the method shown in~\cite{bartnick2021cryogenic}. By performing a step-wise cool-down of waveguides with five different poling periods and observing the shift in the phase-matched SHG wavelength, we can add an empirical correction to the phase-matching function to more accurately describe the change in the spectral properties. This is done by modifying Eq.~(\ref{eqn:mismatch}) with $\Delta n(\lambda_\mathrm{p},T) \rightarrow \Delta n(\lambda_\mathrm{p},T) + \delta n(T)$, where $\delta n(T)$ is a temperature dependent fifth-order polynomial function. 
The measured wavelength shifts for the five waveguides, together with the uncorrected and corrected simulation, are shown in Fig.~\ref{fig:SHGShift}. While both theoretical models describe the wavelengths well around room temperature, the empirical correction is required to provide accurate description of the cryogenic SHG wavelength. 
Using this modification, we choose a poling period of \SI{8.81}{\micro m} to achieve cryogenic degeneracy in the telecom C-band.

 \section{Characterizing cryogenic phase-matching}
We fabricate our waveguide sample including periodic poling by using a laser lithography tool. The waveguide structures are defined by titanium stripes which are indiffused into $z$-cut congruently grown lithium niobate. Afterwards, the electrode patterns for the poling process are written and used to periodically invert the crystal polarization. For the waveguide investigated in this work we chose a poling period of \SI{8.81}{\micro m} and a length of \SI{24.3(2)}{mm}. According to our simulations, we expect the phase-matched wavelength for the chosen poling period to be ${\lambda_{\mathrm{pm,theo}} = \SI{1558.1}{nm}}$ at a temperature of \SI{6.4}{K}, which is the base temperature of our cryostat.
This means that we expect degenerate SPDC photons centered at this wavelength, when the cryogenic process is pumped at ca. \SI{779}{nm}.

\subsection{Cryogenic second harmonic generation measurements}
\begin{figure}[t]
\centering\includegraphics[width=0.90\linewidth]{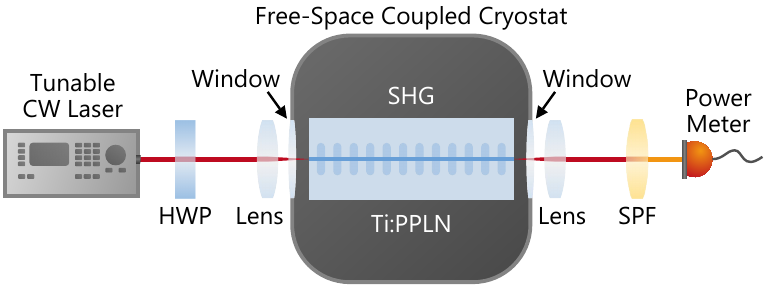}
\caption{\label{fig:SetupSHG}Experimental setup to measure the second harmonic generation (SHG) spectrum of the cryogenic waveguide, placed inside a free-space coupled cryostat. HWP: half-wave plate, Ti:PPLN: titanium-indiffused periodically poled lithium niobate, SPF: short-pass filter.}
\end{figure}

The experimental setup to detect the SHG signal is shown in Fig.~\ref{fig:SetupSHG}. The waveguide is placed inside a free-space coupled cryostat, which provides direct access to the waveguide end facets through transparent windows. The laser beam is coupled to the waveguide using two aspheric lenses, equipped with an anti-reflection coating, placed outside the cryostat. The SHG process is pumped with a continuous wave laser source, that can be tuned in the range from \SI{1440}{nm} to \SI{1640}{nm}. The optical pump power is set to \SI{22.0(01)}{mW}. The laser is followed by a half-wave plate to set the input polarization. The unconverted pump light is filtered by a short-pass filter, and the SHG power is then detected with a power meter. 

\subsection{Analysis of the spectral results}
\begin{figure}[b]
\centering\includegraphics[width=0.88\linewidth]{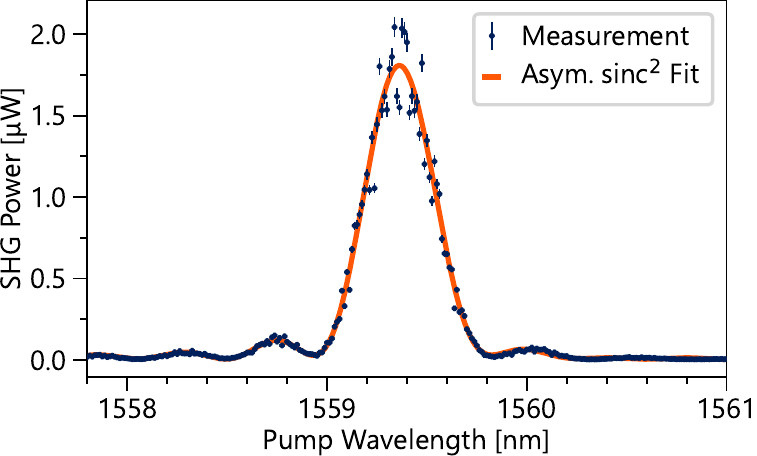}
\caption{\label{fig:SHGSpectrum}Measured data for the cryogenic SHG power over the pump wavelength. The asymmetric fit function considers a non-uniform refractive index distribution along the waveguide direction. The errorbars correspond to an uncertainty of ${\pm \SI{3}{\%}}$, as specified for the used power meter.}
\end{figure}
\begin{figure*}[t]
\centering\includegraphics[width=0.99\textwidth]{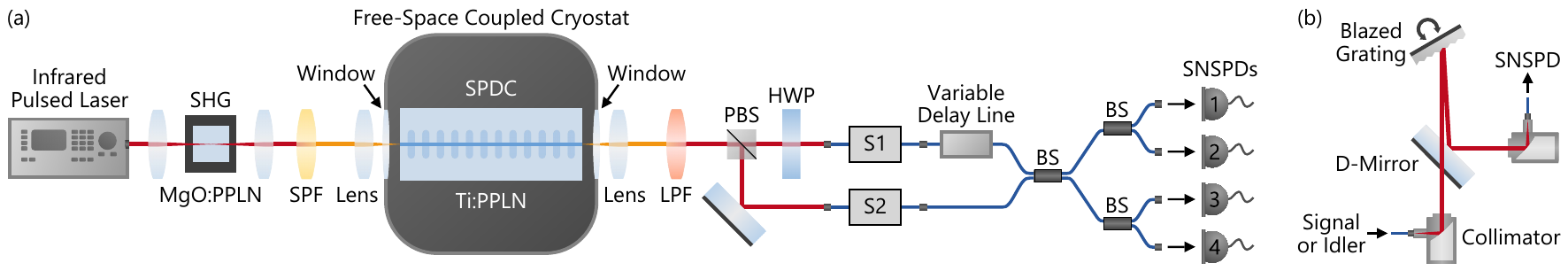}
\caption{\label{fig:SetupHOM}(a) Experimental setup to perform the Hong-Ou-Mandel interference measurement of the cryogenic SPDC source. One scanning-grating spectrometer setup (b) is inserted into each path, in order to spectrally filter signal and idler. MgO:PPLN/Ti:PPLN: MgO-doped/titanium-indiffused periodically poled lithium niobate, SPF/LPF: short-/long-pass filter, (P)BS: (polarizing) beam splitter, HWP: half-wave plate, S1/S2: spectrometer 1/2.}
\end{figure*}
We tune the pump wavelength while measuring the generated SHG power. The measurement data is presented in Fig.~\ref{fig:SHGSpectrum}. According to Eq.~(\ref{eqn:shgPower}), the phase-matching function is expected to show a $\mathrm{sinc}^2$ shape. This description holds for an ideal process, while the actual interaction inside the lithium niobate waveguide can be altered by fabrication imperfections, pyroelectricity and photorefraction~\cite{weis1985lithium}. The pyroelectric and photorefractive effect can lead to an accumulation of electric charges inside the waveguide region. The interplay of those effects can cause localized perturbations of the refractive indices~\cite{bartnick2021cryogenic}, induced by temperature changes during the cool-down~\cite{parravicini2011all}, or high optical power~\cite{rams2000optical}. 

Our measured spectrum is indeed very close to the ideal $\mathrm{sinc}^2$-function, the overlap is \SI{99.9}{\%}. We can more accurately model the exact shape by considering a non-uniform refractive index distribution along the waveguide. This results in a very slight asymmetry of the $\mathrm{sinc}^2$-function, which takes into account imperfections of the phase-matching. For this analysis we follow the procedure presented in~\cite{bartnick2021cryogenic}. The fit reveals an effective length of ${L_{\mathrm{eff}} = \SI{18.8(03)}{mm}}$, corresponding to \SI{77.4(14)}{\%} of the total waveguide length. This means that the phase-matching condition is satisfied for this fraction of our waveguide, which demonstrates precise fabrication of a uniform poling period.

In addition to the effective length of the waveguide, we can extract the phase-matched wavelength ${\lambda_{\mathrm{pm,meas}} = \SI{1559.36(2)}{nm}}$ from the SHG measurement. The error of this value is given by the absolute wavelength accuracy of our laser source. In comparison to the theoretical value, the measured wavelength is slightly higher. Various factors can contribute to this discrepancy. On the one hand, our simulation is based on an empirical refractive index correction term, which induces an experimental uncertainty. On the other hand,  perturbations in the refractive indices can induce an additional shift of the phase-matched wavelength. In fact, the phase-matched wavelength changed to ${\lambda_{\mathrm{pm,meas}} = \SI{1558.60(2)}{nm}}$ in a second cool-down, which was performed more than a month later.
Hence, a slight variation in the phase-matching can arise for every cool-down cycle.
We thus assume spontaneous changes in the accumulating charges to be the primary cause of the discrepancy between experiment and theory. Nevertheless, the experimental wavelengths are indeed very close to our theoretical result, which shows accurate fabrication and good predictability of the phase-matching behavior under cryogenic conditions.

\section{Measuring cryogenic SPDC photons}
Following the SHG characterization, we use the setup shown in Fig.~\ref{fig:SetupHOM}~(a) to pump the SPDC process. We pump the cryogenic SPDC with the SHG signal generated in a bulk periodically poled MgO-doped lithium niobate crystal. This crystal is pumped by a pulsed infrared laser with a repetition rate of \SI{80}{MHz}, a wavelength of \SI{1556.4(2)}{nm} and a bandwidth of \SI{11.81(2)}{nm}. The SHG crystal is temperature controlled to allow fine adjustment of the generated SHG wavelength within a few nanometers. This way, we set the pump wavelength to fit our phase-matched wavelength. We use a $4f$-line to set the spectral bandwidth to \SI{0.73(5)}{nm}. Short-pass filters in front of the cryostat filter out the residual infrared components, before pumping the SPDC chip. The signal and idler photons are filtered by long-pass filters and separated by a polarizing beam splitter.  

We start by investigating the spectral properties with a home-built scanning-grating spectrometer setup shown in Fig.~\ref{fig:SetupHOM}~(b). One spectrometer is placed in each path. The wavelength components of the signal or idler photons are separated by a blazed grating, before part of the spectrum is coupled into a polarization-maintaining single-mode fiber. Further details on the spectrometer setup design can be found in the supplemental document of~\cite{Lange2022cryogenic}. For the spectral measurements, the spectrometers are directly connected to superconducting nanowire single-photon detectors (SNSPDs). The detectors are positioned in a separate cryostat. A time-tagging module is used to record the coincidence counts of the two detectors.

\subsection{Joint spectrum of signal and idler}
\begin{figure}[b]
\centering\includegraphics[width=0.98\linewidth]{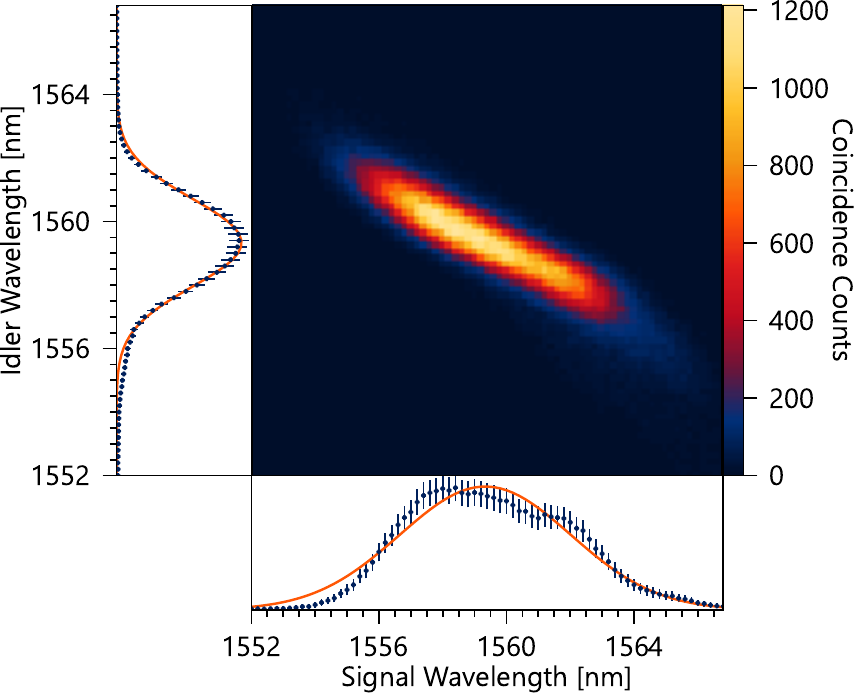}
\caption{\label{fig:JSI}Measured joint spectral intensity (JSI) of the signal and idler photons, together with the marginal spectra. Here, the errorbars correspond to Poisson errors and a Gaussian fit is applied to the marginal data.}
\end{figure}
We measure the joint spectral intensity (JSI) of signal and idler photons by scanning the two gratings and detecting the coincidence counts. For this measurement, we set the optical power of the pump beam to \SI{133(1)}{\micro W} and the integration time for every measurement point to \SI{4}{s}. The measurement data is shown in Fig.~\ref{fig:JSI}. We present the JSI together with the marginal spectra of signal and idler, which correspond to the projection on the two axes. We apply a Gaussian fit to the marginals to obtain the central wavelengths and bandwidths. The shape of the signal spectrum deviates from a Gaussian shape, which is caused by a slight asymmetry of our pump spectrum, after filtering with the $4f$-line. Nevertheless, the fit matches the data well enough to provide the main parameters.
The signal spectrum has a central wavelength of 
${\lambda_\mathrm{s} = \SI{1559.3(5)}{nm}}$ and a bandwidth of ${\Delta\lambda_\mathrm{s} = \SI{6.33(12)}{nm}}$, while the idler spectrum is centered at the same wavelength
${\lambda_\mathrm{i} = \SI{1559.3(6)}{nm}}$ with a smaller bandwidth of ${\Delta\lambda_\mathrm{i} = \SI{3.26(3)}{nm}}$. The uncertainties of the central wavelengths correspond to the transmission bandwidths of the spectrometers, and the errors of the marginal bandwidths arise from the uncertainty in the spectrometer calibration and the error of the Gaussian fit.
For the signal and idler spectra, we obtain a difference in the bandwidths. This is due to the angular orientation of the JSI, which is fundamental to the phase-matching, and thus the material properties of lithium niobate. The spectral purity of our photons is limited by correlations in the signal and idler wavelengths. We analyze these correlations by performing a Schmidt decomposition~\cite{ekert1995entangled}, which results in a Schmidt number of 2.72, corresponding to a spectral purity of 0.37. The purity can be increased by spectrally filtering in the signal and idler arm, as we do for the HOM interference measurement (see Section \ref{sec:hom}).

The measurement of the JSI shows that the signal and idler wavelengths are identical and very close to the measured phase-matched wavelength for SHG. In comparison to the SHG spectrum, the JSI was measured about one week later, during which the waveguide was kept at cryogenic temperatures. This clearly verifies that in spite of the build up of pyroelectric charges, which can disturb the phase-matching during temperature change, the SPDC source allows for stable operation at a constant temperature. 

\subsection{SPDC source performance}
The experimental setup is further modified to measure the source brightness $B$, the Klyshko efficiency $\eta_\mathrm{Klyshko}$~\cite{klyshko1980use}, and the heralded second-order correlation function $g_\mathrm{h}^{(2)}(0)$. For the first two measurements, the spectrometers are removed and the signal and idler photons are each connected to an SNSPD. For the $g_\mathrm{h}^{(2)}(0)$-measurement, a 50:50 fiber beam splitter is added to the idler path to characterize the photon number statistics. The SPDC chip is pumped with an optical power of \SI{66.1(01)}{\micro W}, corresponding to a low photon-pair generation probability of 0.01.

We define the source brightness by ${B = C_\mathrm{si}/P_\mathrm{trans}}$, where $C_\mathrm{si}$ is the signal and idler coincidence rate and $P_\mathrm{trans}$ is the transmitted pump power. Our source has a brightness of ${B = \SI{6.0(03)}{\times10^5~pairs/smW}}$. 
The combined Klyshko efficiency for both paths is given by ${\eta_\mathrm{Klyshko} = \sqrt{C_\mathrm{si}^2/(C_\mathrm{s} C_\mathrm{i})}}$, with $C_\mathrm{s}$ and $C_\mathrm{i}$ the single count rates of signal and idler. We obtain an efficiency of ${\eta_\mathrm{Klyshko} = \SI{13.62(008)}{\%}}$. 
At the same time we measure a low $g_\mathrm{h}^{(2)}(0)$-value of ${g_\mathrm{h}^{(2)}(0) = 0.017 \pm 0.002}$. The heralded autocorrelation function is given by ${g_\mathrm{h}^{(2)}(0) = (C_\mathrm{i_1i_2s} C_\mathrm{s})/(C_\mathrm{i_1s} C_\mathrm{i_2s})}$, where $C_\mathrm{i_1s}$ and $C_\mathrm{i_2s}$ are the coincidences of the two idler paths with the signal photons respectively, and $C_\mathrm{i_1i_2s}$ are the threefold coincidences. Our result for the heralded $g_\mathrm{h}^{(2)}(0)$ is well below the threshold for a two-photon fock state (${g_\mathrm{h}^{(2)}(0) = 0.5}$), which clearly verifies cryogenic single-photon generation.

\subsection{Demonstration of indistinguishability} \label{sec:hom}
After we have shown that our SPDC source generates single photons at cryogenic temperatures and the signal and idler spectra overlap very well, we demonstrate the indistinguishability in the other degrees of freedom. For this purpose, we perform a Hong-Ou-Mandel interference measurement by using the experimental setup as shown in Fig.~\ref{fig:SetupHOM}~(a). The spectrometers S1 and S2 are connected to the signal and idler path, acting as spectral band-pass filters.
This allows us to increase the spectral purity and therefore the indistinguishability of our photon source. Our filters have slightly different transmission bandwidths of ${\Delta \lambda_\mathrm{s} = \SI{0.96(3)}{nm}}$ and ${\Delta \lambda_\mathrm{i} = \SI{1.12(3)}{nm}}$ due to experimental imperfections in the home-built setups. We adjust the spectral filters by setting one filter to the central wavelength of the marginals and fine tune the other grating to maximize the spectral overlap. However,
small differences in the grating position and spectral bandwidth will limit the visibility of our HOM dip.

\begin{figure}[t]
\centering\includegraphics[width=0.88\linewidth]{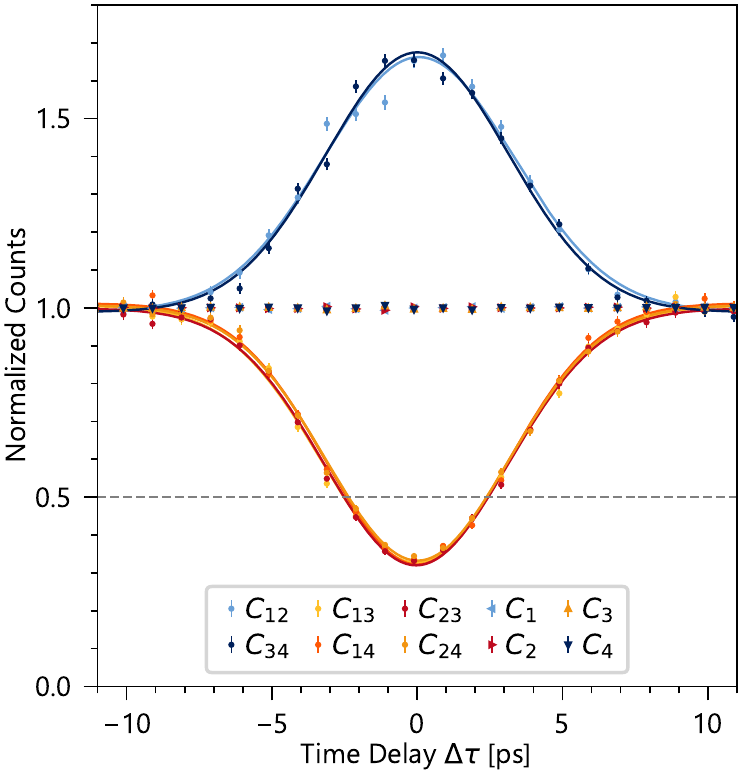}
\caption{\label{fig:HOMDip}Hong-Ou-Mandel interference dip measurement of the cryogenic SPDC photons. The plot shows the normalized count rates for all twofold coincidences (circles) and single counts (triangles) of the SNSPDs shown in Fig.~\ref{fig:SetupHOM}. The errorbars correspond to Poisson errors.}
\end{figure}
We use a type-II SPDC process, which results in orthogonal polarization of signal and idler. Therefore, a half-wave plate is placed in front of one spectrometer to rotate the signal polarization by $90^\circ$. To guarantee stable polarization, all fibers used in this setup are polarization maintaining. Following the spectrometer, we add a fiber-coupled delay line to tune the temporal delay between signal and idler, before both photons meet at the subsequent 50:50 beam splitter. After this interference beam splitter, both paths are split again with another 50:50 beam splitter, before each signal is detected with an SNSPD. We measure the twofold coincidence rates between all four detectors to gain information about the photon statistics in each arm. The temporal delay $\Delta\tau$ is tuned in steps of \SI{1}{ps} and the measurement of the coincidence rates, together with the single count rates, is shown in Fig.~\ref{fig:HOMDip}. 

For perfect temporal overlap (${\Delta\tau = 0}$) we clearly see a peak in the coincidence rates $C_{12}$ and $C_{34}$.  
Simultaneously, there is a significant dip in the twofold coincidences between the different interference paths. The dip goes below the threshold of $0.5$, which verifies bunching of the signal and idler photons. We calculate the HOM visibility of the photons from the sum of all coincidence counts which appear between different interference paths ${C = C_{13}+C_{14}+C_{23}+C_{24}}$. The visibility is thus defined by ${V = 1-C(\Delta\tau = 0)/C(\Delta\tau \rightarrow \infty)}$, which gives us a visibility of ${V = \SI{66.3(5)}{\%}}$. This visibility is limited by imperfections in the spectral filtering, the spatial mode overlap, adjustment of photon polarization, and performance limits of the polarizing beam splitter.
Nevertheless, we can clearly show the bunching of signal and idler, and
therefore, that we can obtain degenerate photons from our cryogenic single-photon source.

\section{Conclusion}
%%% CONCLUSION %%%
Cryogenic SPDC is a crucial component for integrated quantum optics with superconducting detectors.
Our Hong-Ou-Mandel measurement shows that we have successfully generated degenerate photon pairs in the telecom C-band from a cryogenic integrated SPDC source. When operated under ambient conditions, modifying the phase-matching using temperature tuning is a common method to achieve degeneracy, but temperature tuning cannot be used for cryogenic operation.  
We have demonstrated a precise theoretical model to predict the poling period which is required to achieve efficient phase-matching for the cryogenic degenerate process. This is indispensable since the periodic poling is the only parameter we can use to tune the phase-matching in a cryostat and it needs to be defined before fabrication. Despite the fact that no absolute values for the cryogenic effective refractive indices are available, and that pyroelectricity and photorefraction can locally perturb the phase-matching, our experimental wavelength is within \SI{1.5}{nm} of our design. Moreover, our phase-matching function is very close to the ideal $\mathrm{sinc}^2$ shape. We thus show accurate and uniform fabrication of the poling period, maintained under cryogenic conditions, which is significant to realize efficient and customized frequency conversion devices for future cryogenic integration. Our results demonstrate reliable control of the cryogenic nonlinear interaction, which will help to realize fully integrated cryogenic quantum circuits to promote the proliferation of quantum technologies.

\begin{acknowledgments}
We acknowledge financial support from the Deutsche Forschungsgemeinschaft (231447078–TRR 142) and the Bundesministerium für Bildung und Forschung (13N14911). Co-funded by the European Union (ERC, QuESADILLA, 101042399). Views and opinions expressed are however those of the author(s) only and do not necessarily reflect those of the European Union or the European Research Council. Neither the European Union nor the granting authority can be held responsible for them.
\end{acknowledgments}

% \appendix
% \section{Appendixes}

\bibliography{library}% Produces the bibliography via BibTeX.

\end{document}